# GENERAL RELATIVITY AND SPATIAL FLOWS:
# I.  ABSOLUTE RELATIVISTIC DYNAMICS[*]

Tom Martin
Gravity Research Institute
Boulder, Colorado 80306-1258
tmartin@rmi.net

## Abstract

*Two complementary and equally important approaches to relativistic physics are explained.  One is the standard approach, and the other is based on a study of the flows of an underlying physical substratum.  Previous results concerning the substratum flow approach are reviewed, expanded, and more closely related to the formalism of General Relativity.  An absolute relativistic dynamics is derived in which energy and momentum take on absolute significance with respect to the substratum.  Possible new effects on satellites are described.*

## 1.  Introduction

There are two fundamentally different ways to approach relativistic physics.  The first approach, which was Einstein's way [1], and which is the standard way it has been practiced in modern times, recognizes the *measurement reality* of the impossibility of detecting the absolute translational motion of physical systems through the underlying physical substratum and the *measurement reality* of the limitations imposed by the finite speed of light with respect to clock synchronization procedures.  The second approach, which was Lorentz's way [2] (at least for Special Relativity), recognizes the *conceptual superiority* of retaining the physical substratum as an important element of the physical theory and of using *conceptually useful* frames of reference for the understanding of underlying physical principles.

Whether one does relativistic physics the Einsteinian way or the Lorentzian way really depends on one's motives.  The Einsteinian approach is primarily concerned with

---

being able to carry out practical space-time experiments and to relate the results of these experiments among variously moving observers in as efficient and uncomplicated manner as possible. For this reason, the Einsteinian approach is primarily based on an epistemology which is concerned with *practical measurement strategies*. In the restricted case of Special Relativity (inertial frames of reference), it provides us with the simplification of our not having to contend with the underlying physical substratum (absolute space in mechanics or the aether in electrodynamics) *at all*. However, it pays for this simplification by being somewhat physically incomprehensible and by being prone to all the misunderstandings, the confusion, and the seeming paradoxes which inevitably arise from the Einstein synchronization process (clock synchronization by light signals and the *measurement strategy assumption* of isotropic constancy for the speed of light in all inertial frames). The inertial coordinate frames of Special Relativity and, to a first approximation, the Schwarzschild-like coordinate frames of General Relativity are all based on these Einstein synchronization procedures. They are practically realizable and can be set up by using simple and familiar radar and time transfer techniques.

On the other hand, the Lorentzian approach to space-time physics is based on the *conceptual advantages* of retaining an underlying physical substratum in physical theories (which implies anisotropy for the speed of light) and of using Galilean frames of reference among which there exists absolute time (equivalently, absolute simultaneity) to *envision* and *understand* the characteristic phenomena of the theories. These Galilean frames of reference are useful for our *imaginations* and for our *conceptual understanding* of the physics, but they are usually very difficult to set up as practically functioning physical frames of reference. This difficulty arises, because the achievement of absolute simultaneity between reference frames essentially requires that we synchronize the coordinate clocks with infinitely fast signals (which are obviously not available). But why should these practical difficulties lead us to limit our *mental understanding* of the essentials of physical phenomena by restricting our *thought processes* only to frames of reference which are easily achievable with the usual limited scientific instrumentation? This would be as ill-advised in the conceptual arena as to attempt to always set up Galilean frames of reference in which to carry out our experiments and relate our results to other observers (who would also be struggling to set up such frames) in the measurement arena. We shall discover that there is much to be gained by allowing our minds to roam beyond the epistemological limitations of our scientific instruments.

We are suggesting that the Einsteinian and Lorentzian views of space-time physics complement each other. The Einsteinian view is best used for setting up simplified and practical *measurement schemes* (we have a great body of experimental evidence for the success of this approach in the past century), while the Lorentzian view may prove to be the best way to achieve a *conceptual understanding* of the underlying physics and possibly *new physical insights* (we are especially advocating a renewal of this method to *gravitational physics*).

As a simple example of this complementary relationship which has been evident to some well-known physicists [3,4], we mention that it is always possible in Special Relativity to make a continuous transformation from the usual inertial coordinates in the Einsteinian view to Galilean coordinates in the Lorentzian view in which the physical substratum exists and has a temporally constant and spatially homogeneous flow. With the use of these Galilean coordinates and the reification of the substratum, all of the seemingly mystifying aspects of Special Relativity in the Einsteinian view are swept away. But, we can take this a step further. Since the Principle of Covariance holds in General Relativity, it also holds in Special Relativity. Thus, the continuous transformation relating the Einstein synchronized inertial coordinates and the Galilean coordinates assures us that no physical experiment (one carried out with ordinary laboratory instruments) can distinguish between the Einsteinian view and the Lorentzian view in Special Relativity.

Historically, Lorentz usually thought of the underlying physical substratum as being the temporally constant and spatially homogeneous substratum of Newton's absolute and immutable space. In contrast to Lorentz and Newton, we are very much interested in *generalizing* the substratum view to completely arbitrary spatial flows. We envision the possibility that physical space is flowing into or out of planetary bodies, that it is swirling and spiraling in galactic structures, and that in certain circumstances, its flow might even be superluminal.

In this paper, we will review and expand some of the implications of the results obtained in previous work [5,6], relate them a little more closely to the formalism of General Relativity, and sketch the rudiments of an absolute relativistic dynamics (in which energy and momentum take on an absolute significance with respect to the substratum). Our understanding of this dynamics will be enhanced by using the physically intuitive notation and techniques of ordinary vector analysis [13,14]. We are primarily interested in determining how physical space transfers energy and momentum to material bodies as they move through it in forced or unforced motion. We will also review a few of the implications of this spatial flow dynamics for possibly new effects in the celestial mechanics of satellites.

A word about terminology: we have been referring to the flowing physical substratum as the flow of physical space (or more succinctly as the flow of space). As we mentioned above, in the guise of Newtonian mechanics, this substratum has always been referred to as absolute space. In the electromagnetic realm, it has commonly been called the aether. In more modern times, it has even been associated with the rather vague concept of a quantum vacuum. The reader is free to choose whatever terminology he desires, as long as he does not inadvertently bring along any of its historical baggage. For our purposes, the flow of physical space is completely characterized by a 3-space vector field $\mathbf{w} \equiv \mathbf{w}(\mathbf{r}, t)$ in a global Galilean coordinate frame $\{\mathbf{r}, t\}$ on the space-time manifold. The only other defining feature is that the speed of light with respect to this substratum is isotropic and equal to the usual constant $c$.

## 2. Relativity and Spatial Flows

The papers [5] and [6] presented the *essential steps* one must take in order to discuss the possibility that all of relativistic physics (gravitation included) can be understood in terms of the interactions of physical systems with a flowing underlying physical substratum (physical space). This interpretation, with its enhancement of physical understanding, was already available in the restricted case of Special Relativity as mentioned above (see [2,3,4]). The papers [5] and [6] suggest the means by which this enhanced physical understanding might be realized in General Relativity as well.

Absolute rotational motion with respect to physical space has been known since ancient times (the ordinary experience of centrifugal forces). The first optical experiment detecting absolute rotation was performed by Sagnac and reported in 1913 [7]. Laser gyros are used routinely in the present era to detect the smallest rotations for navigational purposes. In these *rotational* experiences, we are made aware of the physical reality of the underlying substratum.

In contrast to the rotational case, Nature is very subtle and illusive when it comes to our detecting any translational motion through physical space. One might say that this is *the paramount fact* of all of physics. It was certainly an important ingredient in the evolution of Einstein's approach to relativistic physics (physical indistinguishability of all inertial frames), and it involved the minds and monumental efforts of some of the greatest experimental physicists in the seemingly fruitless attempt to detect it. The null results of their experiments forced physicists to look closer at the interrelationship of the substratum flow, electrodynamics, and the structure of matter. It became apparent that their experiments would be *intrinsically* insensitive to the translational flow they were attempting to detect.

We should not be disturbed by this inherent design of Nature, because, in a deeper physical sense, it is the very essence of our freedom to travel through unlimited space (Galileo's Principle).

It is doubtful, therefore, that translational motion through physical space will ever be detected by a *localized* laboratory experiment. Nevertheless, the papers [5] and [6] have some *new nonlocalized experiments* to suggest which might be effective in exposing the physical reality of the translational motion of physical space. These experiments are based on the possibility that Nature might not be so successful at hiding the reality of the flow at the *boundary* between two separate flows. In [5], we reminded the reader of the fact that there are *spatial inflow and outflow solutions* of Einstein's field equations for planetary objects in General Relativity which might have physical reality. These flowing space solutions are actually *a part of* and actually *predicted by* General Relativity. If we are correct in educing that such flows of space will actually be along the gravitational accelerative field lines for slowly moving test bodies (Bernoulli analog), then, in the two-body gravitational problem, there will be a *boundary* separating the flows of each of the two bodies. We are not yet precisely certain of the exact configuration the boundary will take in the case of two mutually orbiting bodies, but there will be a stagnation point for the two flows somewhere in the larger neighborhood of the gravitational saddle point of the Newtonian gravitational potentials. In [5], we described a simple satellite experiment which can detect this region of stagnation with surprising sensitivity. This might be the first experiment which establishes the *physical reality* of the translational flow of space. Then, in [6], we revealed that there might be dynamical effects on satellites (as well as time dilation jumps) as they pass through the two-body boundary. These dynamical effects might be the first to reveal the actual *direction* of the translational flow.

There are two reasons why these boundary effects on satellites have not yet been consciously observed (if, in fact, the flow solutions correspond to physical reality). The first reason is that these effects have not been expected, and if they have been observed, they will have most probably been interpreted as spurious affects. Secondly, most satellites are tracked by the Doppler radar technique. An uplink signal is received by a transponder onboard the satellite, and the downlink signal is *phase-locked* to this uplink signal so the transponder effectively acts as a perfect mirror (with gain). This phase-locking makes the signals insensitive to any time dilation affects which might be experienced by the internal oscillators onboard the satellite. A *freely running* and sensitive clock must be onboard the satellite in order to detect the hypothetical time dilation jumps at the two-body boundary.

Whether or not the spatial flow solution for the two-body problem turns out to correspond to physical reality, we know from the Principle of Covariance that no physical experiment can distinguish the Galilean flow solution from the curved space Schwarzschild solution in the case of a single *isolated* attractor (the near field of the Earth is approximately that of an isolated attractor). We can always use the flow and its associated Galilean coordinate frame as a perfectly valid and accurate conceptual model with which to study the exact details of the effects of such an attractor on physical systems. Hence, it is probably useful to mention the following elementary comparison between the nonstatic Galilean coordinates and the static and spatially curved Schwarzschild coordinates of such an attractor (see reference [5] for the mathematical details).

When the space-time manifold associated with a simple gravitational attractor is coordinated with Schwarzschild coordinates, the coordination implies that nonmoving meter rods laid out in the radial direction are shrunk, while those oriented in the horizontal position are not. The shrinking in the radial direction is interpreted as being "caused" by the non-Euclidean nature of space. Because of this interpretation, it is necessary to analyze everything from a differential geometric point of view and methodology.

When this very same space-time manifold is coordinated with Galilean coordinates, nonmoving meter rods laid out in the radial direction are shrunk by the Lorentz contraction arising from the inflowing or outflowing substratum. Nonmoving meter rods in the horizontal orientation are not shrunk, because there is no flow through them in their lineal direction. This gives rise to the same "non-Euclidean geometry" (from the viewpoint of placing the meter rods end to end), but we get to talk about it in conceptual terms as being "caused" by the flow of physical space (in the Euclidean coordinate space of the Galilean frame). We can talk about it without necessarily bringing in all the machinery of differential geometry. This was one of the purposes which motivated the writing of reference [5]. For a spherically symmetric gravitational attractor, we can get all of the physical predictions of General Relativity without having to do differential geometry *per se*. Reference [6] was similarly motivated, and it shows us how to determine the trajectories of test bodies in completely general spatial flows without having to use the mathematical machinery of differential geometry *per se* (no covariant differentiation, no Christoffel symbols, no affine parameters, and so forth). So, the spatial flow point of view has the potential to bring tremendous conceptual and mathematical simplification to all of space-time physics and especially to General Relativity.

Since there is a possibility that the gravitational field of the Earth is actually caused by an inflow or outflow of the physical substratum, we are naturally led to wonder about what could be the explanation of *stellar aberration* when there is vertical substratum flow. Without going into any detail here, we simply state that, if the substratum is flowing essentially vertically into or out of planetary objects, we must use Stokes' 1845 explanation of stellar aberration [8,9]. Lorentz's objection [10] to Stokes' explanation was misplaced. Lorentz assumed (as did most physicists of the day) that, if the substratum was entrained at the Earth's surface, it would be *horizontally* entrained. The curl of the actual *vertical* flow is zero, and so it meets the requirements of Stokes' explanation.

Finally, as an historical curiosity, we mention that we know today that Michelson and Morley's famous optical interferometer [11] was inherently insensitive to spatial flow of any kind (because of the electrodynamic phenomenon of Lorentz contraction). Still, there is a bit of irony in the possibility that the substratum was amiably flowing *vertically* through their interferometer as they patiently rotated it in the horizontal plane. Michelson, however, was the one who had the last word. He later measured the small horizontal component in the vertical substratum flow (which arises in an Earth co-spinning frame) in the superb Michelson-Gale experiment [12].

# 3. Absolute Relativistic Dynamics

We now proceed to establish the rudiments of relativistic dynamics for test bodies in arbitrary spatial flows. If the spatial flow approach to space-time physics is universally applicable to all physical circumstances, every space-time manifold of physical significance will be capable of being characterized by a flow of physical space. A global Galilean coordinate frame $\{\mathbf{r}, t\}$ will always exist in which the flow is represented by a global 3-space vector field $\mathbf{w} \equiv \mathbf{w}(\mathbf{r}, t)$. The rate of an atomic clock will depend on its *absolute speed* $u$ with respect to physical space. Thus, in these Galilean coordinates, the proper time element of an atomic clock will be given by

$$d\boldsymbol{t} = \boldsymbol{g}^{-1} dt \equiv \sqrt{1 - u^2 / c^2} \, dt = \sqrt{1 - \mathbf{u} \cdot \mathbf{u} / c^2} \, dt \quad , \tag{3-1}$$

where

$$\mathbf{u} \equiv \mathbf{v} - \mathbf{w} \tag{3-2}$$

is the *absolute velocity* of the clock relative to physical space. Here, $\boldsymbol{t}$ is the proper time of the clock, $c$ is the speed of light with respect to physical space (a constant), $t$ is the coordinate time, $\mathbf{r}$ is the coordinate position vector of the clock, and $\mathbf{v} \equiv d\mathbf{r} / dt$ is the coordinate velocity of the clock.

From the above equations, we see that the space-time line element in Galilean coordinates with arbitrary spatial flow $\mathbf{w}$ always takes the form

$$c^2 d\boldsymbol{t}^2 = (c^2 - w^2) dt^2 + 2\mathbf{w} \cdot d\mathbf{r} \, dt - (d\mathbf{r})^2 . \tag{3-3}$$

In four dimensional *rectangular* Galilean coordinates $\{x^0, x^1, x^2, x^3\} \equiv \{ct, x, y, z\}$, the components of the corresponding space-time *metric tensor* are seen to be

$$g_{kl} = \begin{bmatrix} -(1 - w^2 / c^2) & -w^x / c & -w^y / c & -w^z / c \\ -w^x / c & 1 & 0 & 0 \\ -w^y / c & 0 & 1 & 0 \\ -w^z / c & 0 & 0 & 1 \end{bmatrix} , \tag{3-4}$$

where we use the convention that

$$ds^2 \equiv c^2 d\boldsymbol{t}^2 \equiv -g_{kl}\, dx^k\, dx^l \quad . \tag{3-5}$$

From (3-3) or (3-4), it is immediately evident that every slice of space-time with constant time ($dt \equiv 0$) is flat 3-dimensional Euclidean space. We also realize that the universality of the spatial flow approach to General Relativity corresponds to the possibility of always being able to impose algebraic *coordinate conditions* in the guise of the special form of the metric tensor in (3-3) and (3-4).

From (3-1) and (3-2), it is evident that the total time dilation of an atomic clock in motion with velocity $\mathbf{v}$ in an arbitrary spatial flow $\mathbf{w}$ is given by

$$d\boldsymbol{t} = \sqrt{1 - (v^2 + w^2 - 2\mathbf{v} \cdot \mathbf{w})/c^2}\, dt \quad . \tag{3-6}$$

In our substratum picture, $c$ is the limiting speed of all material test bodies with respect to *absolute* physical space. Thus, for material test bodies, we will always have the condition $u < c$.

The *absolute relativistic momentum* of a particle with constant restmass $m$ and absolute velocity $\mathbf{u} \equiv \mathbf{v} - \mathbf{w}$ is

$$\mathbf{p} \equiv \boldsymbol{g}\, m\, \mathbf{u} \quad , \tag{3-7}$$

where our $\boldsymbol{g}$ is always the $\boldsymbol{g}$ associated with *absolute motion* in the substratum:

$$\boldsymbol{g} \equiv 1/\sqrt{1 - u^2/c^2} \quad . \tag{3-8}$$

The *total absolute relativistic energy* of such a particle is

$$E \equiv \boldsymbol{g}\, m\, c^2 \quad . \tag{3-9}$$

Since the *restmass energy* is $m\, c^2$, the *absolute relativistic kinetic energy* is

$$K \equiv (\boldsymbol{g} - 1)\, m\, c^2 \quad . \tag{3-10}$$

We know that these are the correct definitions, because an infinitesimal Galilean coordinate frame *comoving* with the substratum is the same as an inertial frame, and these are the correct definitions in the inertial frames of Special Relativity. One can easily verify that these absolute dynamical variables satisfy the fundamental relationship,

$$E^2 = p^2 c^2 + m^2 c^4 \quad . \tag{3-11}$$

The *absolute force* $\mathbf{F}$ acting on the particle is the time rate of change of the absolute momentum:

$$\mathbf{F} \equiv \frac{d\mathbf{p}}{dt} = \frac{d(\boldsymbol{g}\,m\,\mathbf{u})}{dt} = \frac{d\boldsymbol{g}}{dt}m\,\mathbf{u} + \boldsymbol{g}\,m\frac{d\mathbf{u}}{dt} \quad . \tag{3-12}$$

Since $d\boldsymbol{g}/dt = d\big((1-u^2/c^2)^{-1/2}\big)\big/dt = -(1/2)(1-u^2/c^2)^{-3/2}(-2\mathbf{u}/c^2)\cdot d\mathbf{u}/dt$, we have, in complete generality, that

$$\frac{d\boldsymbol{g}}{dt} = \frac{\boldsymbol{g}^3}{c^2}\,\mathbf{u}\cdot\frac{d\mathbf{u}}{dt} \quad . \tag{3-13}$$

Thus, (3-12) becomes

$$\mathbf{F} \equiv \frac{d\mathbf{p}}{dt} = \boldsymbol{g}\,m\frac{\boldsymbol{g}^2}{c^2}(\mathbf{u}\cdot\frac{d\mathbf{u}}{dt})\mathbf{u} + \boldsymbol{g}\,m\frac{d\mathbf{u}}{dt} = \boldsymbol{g}\,m\,(1 + \frac{\boldsymbol{g}^2}{c^2}\mathbf{u}\mathbf{u})\cdot\frac{d\mathbf{u}}{dt} \quad , \tag{3-14}$$

where $1$ is the identity tensor in Galilean 3-space (we are using ordinary vector-dyadic notation and techniques [13, 14]).

The tensor $(1 + (\boldsymbol{g}^2/c^2)\mathbf{u}\mathbf{u})$ is non-singular, and it has the somewhat simpler inverse

$$(1 + (\boldsymbol{g}^2/c^2)\mathbf{u}\mathbf{u})^{-1} \;=\; (1 - \mathbf{u}\mathbf{u}/c^2) \quad . \tag{3-15}$$

*proof:* $(1 + (\boldsymbol{g}^2/c^2)\mathbf{u}\mathbf{u})\cdot(1 - \mathbf{u}\mathbf{u}/c^2) = 1 - \mathbf{u}\mathbf{u}/c^2 + \boldsymbol{g}^2(1-u^2/c^2)\mathbf{u}\mathbf{u}/c^2 = 1$ .

We call the tensor $(\mathbf{1} - \mathbf{uu}/c^2)$ the *tensor of motion* of the particle. Not only does every simple test particle have an associated kinematic velocity vector, it also has an associated kinematic tensor.

From (3-14), we have, in complete generality, that

$$\boxed{\mathbf{F} \;\equiv\; \frac{d\mathbf{p}}{dt} \;=\; \boldsymbol{g}\, m (\mathbf{1} - \mathbf{uu}/c^2)^{-1} \cdot \frac{d\mathbf{u}}{dt}} \;. \tag{3-16}$$

This fundamental relation between the time rate of change of the absolute relativistic momentum (the force) and the *absolute acceleration* $d\mathbf{u}/dt$ of the particle is the generalization of Newton's famous Second Law, $\mathbf{F} = m\,\mathbf{a}$, in arbitrarily flowing absolute space. The equation is *fully relativistic* (in the sense that the particle may have any physically allowable speed ($0 \le u < c$) with respect to physical space).

The fundamental equation (3-16) also shows us that, for relativistic speeds, the *inertia* of the particle is no longer a simple scalar $m$. Instead, the inertia of the particle is a *tensor*. In analogy with the usual form of the Second Law, $\mathbf{F} = m\,\mathbf{a}$, we call the tensor $\boldsymbol{g}\, m (\mathbf{1} - \mathbf{uu}/c^2)^{-1}$ the *inertia tensor* of the particle. Thus, the inertia of a relativistic particle is directionally dependent, and the force and the associated acceleration are no longer necessarily acting in the same direction. Writing (3-16) in the form

$$\frac{d\mathbf{u}}{dt} \;=\; \frac{1}{\boldsymbol{g}\, m}(\mathbf{1} - \mathbf{uu}/c^2) \cdot \mathbf{F} \;=\; \frac{1}{\boldsymbol{g}\, m}\mathbf{F} \;-\; \frac{1}{\boldsymbol{g}\, mc^2}(\mathbf{F} \cdot \mathbf{u})\mathbf{u} \;, \tag{3-17}$$

we see that there is a velocity independent component of the acceleration in the direction of the force as well as a velocity *dependent* component of the acceleration in the direction of the *velocity*. If the force is perpendicular to the particle's motion relative to physical space, (3-17) shows us that the relative strength of its inertia is

$$\textit{transverse inertia} \;\equiv\; \boldsymbol{g}\, m \quad. \tag{3-18}$$

On the other hand, if the force $\mathbf{F}$ is parallel to the velocity, so that $\mathbf{u} = \pm |\mathbf{u}|\mathbf{F}/|\mathbf{F}|$, then $d\mathbf{u}/dt = (\mathbf{F}/\boldsymbol{g}\, m)(1 - u^2/c^2) = \mathbf{F}/\boldsymbol{g}^3\, m$, and the relative strength of its inertia is

$$\textit{longitudinal inertia} \;\equiv\; \boldsymbol{g}^3\, m\,. \tag{3-19}$$

The time rate of change of the total absolute relativistic energy can be calculated from (3-9) and (3-13).  We have, in complete generality, that

$$\boxed{\frac{dE}{dt} \;=\; \boldsymbol{g}^3 \, m\mathbf{u}\cdot\frac{d\mathbf{u}}{dt}} \quad . \tag{3-20}$$

Clearly, the energy of the test particle is constant when the absolute acceleration is orthogonal to the absolute velocity.  No work is done in accelerating the particle *transversely* as it moves through the substratum.  This is why only the longitudinal inertia appears.

By taking the dot product of equation (3-17) with $\mathbf{u}$, one can easily establish that

$$\frac{dE}{dt} \;=\; \mathbf{F}\cdot\mathbf{u} \quad , \tag{3-21}$$

as expected.  The *absolute work*  $dW$  done on the particle by a force $\mathbf{F}$ acting through the absolute substratum distance $d\mathbf{s} \equiv \mathbf{u}\,dt$ is obviously

$$dW \;=\; \mathbf{F}\cdot d\mathbf{s} \;=\; \mathbf{F}\cdot\mathbf{u}\,dt \quad . \tag{3-22}$$

## 4.  The Geodesic Equations of Motion

The path of a *free* material test particle in General Relativity is one over which the integral of the particle's proper time is an extremum:

$$\boldsymbol{d}\int d\boldsymbol{t} = 0 \quad . \tag{4-1}$$

The path is obtained from the associated Euler-Lagrange equations which arise in deriving the solution to the variational problem (4-1).  In the fully generalized differential geometric approach of General Relativity, the equivalent invariant

$$ds \equiv cd\boldsymbol{t} \;\equiv\; \sqrt{-\,g_{kl}\,dx^k\,dx^l} \tag{4-2}$$

is varied, and the resulting Euler-Lagrange equations are the *geodesic equations of motion*,

$$\frac{d^2 x^k}{dt^2} + \Gamma_{lm}^k \frac{dx^l}{dt} \frac{dx^m}{dt} = 0 \quad , \tag{4-3}$$

if one uses an affine parameter such as the proper time $t$. If one uses a non-affine parameter such as coordinate time $t = x^0 / c$, the geodesic equations take on a more complicated form:

$$\frac{d^2 x^k}{dt^2} + \Gamma_{lm}^k \frac{dx^l}{dt} \frac{dx^m}{dt} = -\frac{dx^k}{dt} \frac{d^2 t}{dt^2} \Big/ (\frac{dt}{dt})^2 \quad . \tag{4-4}$$

This differential geometric approach brings in all the mathematical complications of the metric tensor $g_{kl}$, the Christoffel symbols $\Gamma_{lm}^k$, the concept of an affine parameter, and the necessary techniques of generalized tensor analysis.

It is of interest, therefore, to discover that when we are dealing with a space-time manifold which is characterized by the flow of physical space, we have no need to bring in all this generalized mathematical machinery. The equations of motion of free test particles (equivalently, the geodesic equations) can be easily determined within the context of ordinary vector analysis. This was originally demonstrated in reference [6], and because the derivation is so short and simple (and also for the sake of completeness), we repeat it here. Naturally, the calculation is carried out in Galilean coordinates. The calculation is simplified by using the coordinate time $t$ as the path parameter. This reduces the number of Euler-Lagrange equations from four (involving the space-time coordinates) to three (involving the spatial coordinates). From (3-1), we have

$$d \int dt = d \int c^{-1} (c^2 - u^2)^{1/2} dt = 0 \quad . \tag{4-5}$$

We choose *rectangular* Galilean coordinates $\{x, y, z, t\}$ with the orthonormal spatial basis $\{\hat{\mathbf{e}}_x, \hat{\mathbf{e}}_y, \hat{\mathbf{e}}_z\}$. The usual Euler-Lagrange equations are

$$\frac{d}{dt} \left( \frac{\partial L}{\partial \mathbf{v}} \right) = \frac{\partial L}{\partial \mathbf{r}} \quad , \tag{4-6}$$

where

$$L = L(\mathbf{r}, \mathbf{v}; t) = c^{-1}(c^2 - \mathbf{u} \cdot \mathbf{u})^{1/2} = \boldsymbol{g}^{-1} \qquad (4\text{-}7)$$

is the Lagrangian, and

$$\frac{\partial L}{\partial \mathbf{v}} \equiv \frac{\partial L}{\partial v^x} \hat{\mathbf{e}}_x + \frac{\partial L}{\partial v^y} \hat{\mathbf{e}}_y + \frac{\partial L}{\partial v^z} \hat{\mathbf{e}}_z \quad , \qquad (4\text{-}8)$$

$$\frac{\partial L}{\partial \mathbf{r}} \equiv \partial_x L \hat{\mathbf{e}}_x + \partial_y L \hat{\mathbf{e}}_y + \partial_z L \hat{\mathbf{e}}_z \equiv \mathbf{grad}\, L \qquad . \qquad (4\text{-}9)$$

In this Euler-Lagrange formalism, $\mathbf{r}$ and $\mathbf{v}$ are treated as independent variables. $\mathbf{v} \equiv \mathbf{v}(t)$ is a parametric vector, while $\mathbf{w} \equiv \mathbf{w}(\mathbf{r}, t)$ is a vector field. Thus,

$$\mathbf{grad}\, \mathbf{u} = \mathbf{grad}\,(\mathbf{v} - \mathbf{w}) = -\mathbf{grad}\, \mathbf{w} \quad , \qquad (4\text{-}10)$$

where

$$\mathbf{grad}\, \mathbf{w} \;\equiv\; \begin{array}{lll} +\partial_x w^x\, \hat{\mathbf{e}}_x \hat{\mathbf{e}}_x & +\partial_x w^y\, \hat{\mathbf{e}}_x \hat{\mathbf{e}}_y & +\partial_x w^z\, \hat{\mathbf{e}}_x \hat{\mathbf{e}}_z \\ +\partial_y w^x\, \hat{\mathbf{e}}_y \hat{\mathbf{e}}_x & +\partial_y w^y\, \hat{\mathbf{e}}_y \hat{\mathbf{e}}_y & +\partial_y w^z\, \hat{\mathbf{e}}_y \hat{\mathbf{e}}_z \\ +\partial_z w^x\, \hat{\mathbf{e}}_z \hat{\mathbf{e}}_x & +\partial_z w^y\, \hat{\mathbf{e}}_z \hat{\mathbf{e}}_y & +\partial_z w^z\, \hat{\mathbf{e}}_z \hat{\mathbf{e}}_z \end{array} \qquad (4\text{-}11)$$

is the tensor characterizing the *spatial inhomogeneity* of the flow $\mathbf{w}$.

From (4-10), we have

$$-\mathbf{grad}(\mathbf{u} \cdot \mathbf{u}) = -2(\mathbf{grad}\, \mathbf{u}) \cdot \mathbf{u} = 2(\mathbf{grad}\, \mathbf{w}) \cdot \mathbf{u} \;,$$

hence

$$\frac{\partial L}{\partial \mathbf{r}} = \mathbf{grad}\, c^{-1}(c^2 - \mathbf{u} \cdot \mathbf{u})^{1/2} \;=\; \frac{1}{2} L^{-1} c^{-2}\, \mathbf{grad}\,(c^2 - \mathbf{u} \cdot \mathbf{u}) \;=\; L^{-1} c^{-2}\, (\mathbf{grad}\, \mathbf{w}) \cdot \mathbf{u} \;\;.$$

Since

$$L = c^{-1}(c^2 - \mathbf{u}\cdot\mathbf{u})^{1/2} = c^{-1}(c^2 - \mathbf{v}\cdot\mathbf{v} + 2\mathbf{w}\cdot\mathbf{v} - \mathbf{w}\cdot\mathbf{w})^{1/2} \ ,$$

$$\frac{\partial L}{\partial \mathbf{v}} = \frac{1}{2}L^{-1}c^{-2}(2\mathbf{w} - 2\mathbf{v}) = -L^{-1}c^{-2}\mathbf{u} \ ,$$

so

$$\frac{d}{dt}\left(\frac{\partial L}{\partial \mathbf{v}}\right) = -c^{-2}\left(\frac{dL^{-1}}{dt}\right)\mathbf{u} - c^{-2}L^{-1}\frac{d\mathbf{u}}{dt} = c^{-4}L^{-3}\left(\mathbf{u}\cdot\frac{d\mathbf{u}}{dt}\right)\mathbf{u} - c^{-2}L^{-1}\frac{d\mathbf{u}}{dt} \ .$$

Setting $L^{-2} = \boldsymbol{g}^2$ (from (4-7)), we see that our Euler-Lagrange equations (4-6) are

$$\frac{d\mathbf{u}}{dt} + \frac{\boldsymbol{g}^2}{c^2}\left(\mathbf{u}\cdot\frac{d\mathbf{u}}{dt}\right)\mathbf{u} = -(\mathbf{grad}\,\mathbf{w})\cdot\mathbf{u} \ . \tag{4-12}$$

Taking the scalar product of this equation with $\mathbf{u}$, we get

$$\frac{d\mathbf{u}}{dt}\cdot\mathbf{u} + \frac{\boldsymbol{g}^2}{c^2}\left(\mathbf{u}\cdot\frac{d\mathbf{u}}{dt}\right)(\mathbf{u}\cdot\mathbf{u}) = -((\mathbf{grad}\,\mathbf{w})\cdot\mathbf{u})\cdot\mathbf{u} \ . \tag{4-13}$$

Since $\mathbf{u}\cdot\mathbf{u} = c^2(1 - \boldsymbol{g}^{-2})$, (4-13) becomes

$$\boldsymbol{g}^2\,\frac{d\mathbf{u}}{dt}\cdot\mathbf{u} = -((\mathbf{grad}\,\mathbf{w})\cdot\mathbf{u})\cdot\mathbf{u} \ . \tag{4-14}$$

Substituting this back into (4-12), we find that

$$\frac{d\mathbf{u}}{dt} = -(\mathbf{grad}\,\mathbf{w})\cdot\mathbf{u} + \frac{1}{c^2}(((\mathbf{grad}\,\mathbf{w})\cdot\mathbf{u})\cdot\mathbf{u})\,\mathbf{u} \ , \tag{4-15}$$

and we can write this more formally as

$$\boxed{\frac{d\mathbf{u}}{dt} + (1 - \mathbf{u}\mathbf{u}/c^2)\cdot(\mathbf{grad}\,\mathbf{w})\cdot\mathbf{u} = \mathbf{0}} \ . \tag{4-16}$$

Again, $1$ is the identity tensor in 3-space, and $1 - \mathbf{uu}/c^2$ is the symmetric *tensor of motion* of the test particle (this has already arisen in our sketch of relativistic dynamics in Section 3).

(4-16) is the equation of motion of a *free* material test particle in Galilean coordinates with arbitrary spatial flow $\mathbf{w}$. It *is* the geodesic equation in these coordinates. The equation is *fully relativistic* (in the sense that the material test particle may have an arbitrary relativistic speed ($0 \le u < c$) with respect to physical space). Since the Lagrangian (4-7) is Galilean invariant (invariant under the general velocity transformation $\mathbf{q} \mapsto \mathbf{q} + \mathbf{c}$ where $\mathbf{c}$ is a constant vector), the equation itself is Galilean invariant. This is also immediately evident from a perusal of the terms in (4-16). A similar perusal establishes that the equation is time-reversal invariant (invariant under the transformation $t \mapsto -t$).

In contrast to the generalized form of the geodesic equations (4-3) or (4-4), all of the factors appearing in the Galilean form (4-16) are Galilean invariant vectors or tensors.

## 5.  Ponderable and Non-ponderable Forces

We are now in a position to determine how the *free* interactive motion of a test body with the substratum (geodesic motion) transfers absolute momentum and energy to the test body (think of a freely-falling satellite). From equations (3-16) and (4-16), we obtain the time rate of change of the relativistic momentum of a freely-falling test body. We denote this $\mathbf{F}_{substratum}$ :

$$\mathbf{F}_{substratum} \;\equiv\; (\frac{d\mathbf{p}}{dt})_{substratum} \;=\; -\boldsymbol{g}\,m\,(\mathbf{grad}\,\mathbf{w})\cdot\mathbf{u} \quad . \tag{5-1}$$

From equation (3-21), we obtain the time rate of change of its total relativistic energy:

$$(\frac{dE}{dt})_{substratum} \;=\; -\boldsymbol{g}\,m\,((\mathbf{grad}\,\mathbf{w})\cdot\mathbf{u})\cdot\mathbf{u} \quad . \tag{5-2}$$

Since the test body is in free-fall, the force $\mathbf{F}_{substratum}$ is actually a fictitious or *non-ponderable* force acting on the body (incapable of being detected by transducers of any kind within the satellite). Nevertheless, it retains its significance as a time rate of change of momentum. We observe that the transfer of energy and momentum to a freely-falling test body only involves the *motion* $\mathbf{u}$ of the test body and the *inhomogeneities* of the flow (as represented by $\mathbf{grad\,w}$).

In general, the total force acting on a test body is the sum of the *ponderable* (think of rocket thrust) and the *non-ponderable* substratum forces:

$$\mathbf{F} \;\equiv\; \mathbf{F}_{ponderable} \;+\; \mathbf{F}_{substratum} \quad . \tag{5-3}$$

Thus, in general,

$$\mathbf{F}_{ponderable} \;=\; \mathbf{F} - \mathbf{F}_{substratum} \;=\; \boldsymbol{g}\,m\,(1 - \mathbf{uu}/c^2)^{-1} \cdot \frac{d\mathbf{u}}{dt} \;+\; \boldsymbol{g}\,m\,(\mathbf{grad\,w}){\cdot}\mathbf{u} \quad . \tag{5-4}$$

We see that the ponderable force is due to the *temporal acceleration* of the test body $d\mathbf{u}/dt$ with respect to the substratum as well as the motion of the test body through the inhomogeneities of the substratum's flow (*spatial accelerations* of the flow).

_Example 1_: *The gravitational force on a test body on the surface of a planet*.

In spherical Galilean coordinates $\{r, \boldsymbol{q}, \boldsymbol{f}, t\}$ centered on a spherically symmetric attractor, we will first study the gravitational field arising from the substratum *inflow* $\mathbf{w} \equiv -\sqrt{2GM/r}\;\hat{\mathbf{e}}_r$, where $G$ is the gravitational constant and $M$ is the mass of the attractor. This inflow is one of the Galilean representations of the Schwarzschild solution (see reference [5]).

Since the body is at rest on the surface of the planet, $\mathbf{v} \equiv \mathbf{0}$. From $\mathbf{u} \equiv \mathbf{v} - \mathbf{w}$, we have $\mathbf{u} = -\mathbf{w}$ and $d\mathbf{u}/dt = -d\mathbf{w}/dt = -\mathbf{v}{\cdot}\mathbf{grad\,w} - \partial_t\mathbf{w}$ (chain rule) $= \mathbf{0}$ . Thus, equation (3-16) implies $\mathbf{F} = \mathbf{0}$, and equations (5-1) and (5-3) imply

$$\mathbf{F}_{ponderable} \;=\; -\,\mathbf{F}_{substratum} \;=\; \boldsymbol{g}\,m\,(\mathbf{grad\,w}){\cdot}\mathbf{u} \quad . \tag{5-5a}$$

The expression for the tensor $\mathbf{grad\,w}$ in spherical coordinates is commonly available [15]. For the flow $\mathbf{w} \equiv -\sqrt{2GM/r}\;\hat{\mathbf{e}}_r$, we find that

$$\mathbf{grad\,w} = \partial_r \, w^r \, \hat{\mathbf{e}}_r \, \hat{\mathbf{e}}_r = \sqrt{\frac{GM}{2r^3}} \hat{\mathbf{e}}_r \, \hat{\mathbf{e}}_r \quad . \tag{5-5b}$$

Since $\mathbf{u} = -\,\mathbf{w} = \sqrt{2GM / r}\;\hat{\mathbf{e}}_r$, (5-5a) and (5-5b) give us

$$\mathbf{F}_{ponderable} = \boldsymbol{g}\, m \frac{GM}{r^2} \hat{\mathbf{e}}_r \quad . \tag{5-5c}$$

Thus, we are able to understand the experiential force of gravity as we stand on the surface of a planet.  It is a ponderable force directed *upward* on the bottom of our feet. (5-5c), which is correct for relativistic substratum speeds $u$,  shows us that our *weight* is $\boldsymbol{g}\, m$ instead of just $m$.

This ponderable force of gravity (5-5a) and (5-5c) arises, not because the body is temporally accelerated with respect to the substratum, but because it is present in a spatial inhomogeneity of the flow.  Thus, there really is an underlying distinction between the ponderable force of a thrusting rocket and the ponderable force of gravity on the surface of a planet.  The ponderable force of the thrusting rocket does work on its payload, while the ponderable force of gravity on the surface of a planet does no work (the total force $\mathbf{F}$ is zero, and hence, $\mathbf{F} \cdot \mathbf{u}$ is zero).

If we reverse the *direction* of the flow, so that we have the *outflow* $\mathbf{w} \equiv \sqrt{2GM / r}\,\hat{\mathbf{e}}_r$ representing the gravitational field of the spherically symmetric attractor, we see that the signs of $\mathbf{u}$ and $\mathbf{grad\,w}$ will also be reversed.  From (5-5a), we see there will no change in the resulting gravitational force (5-5c).  The inflow and outflow produce the same gravitational effects.  It is not the *direction* of the flow which is responsible for gravitation, but rather, the *inhomogeneities* of the flow.  This is related to the fact that the geodesic equations of motion (4-16) are time-reversal invariant.

*Example 2*:  *A test body in a temporally constant homogeneous substratum flow.*

Here, the flow $\mathbf{w}$ is constant in time and the same at every point of  the Galilean frame.  Thus,

$$\mathbf{grad\,w} = 0 \quad . \tag{5-6a}$$

From (5-1), $\mathbf{F}_{substratum} = \mathbf{0}$, and hence $\mathbf{F}_{ponderable} = \mathbf{F}$. Since $d\mathbf{w}/dt = \mathbf{v} \cdot \mathbf{grad}\,\mathbf{w} + \partial_t \mathbf{w} = \mathbf{0}$ and $d\mathbf{u}/dt = d\mathbf{v}/dt \equiv \mathbf{a}$, equation (3-16) tells us the Galilean acceleration will be

$$\mathbf{a} = \frac{1}{g\,m}(1 - \mathbf{uu}/c^2) \cdot \mathbf{F}_{ponderable} \quad . \tag{5-6b}$$

In the relativistic case, we remember that the ponderable force (rocket thrust) and the resulting acceleration are not necessarily collinear.

$\underline{\textit{Example 3}}$: *A test body on the inner surface of a rotating circular cylinder.*

If the rotating cylinder is rotating with constant rotational velocity $\boldsymbol{\omega}$ around its axis of symmetry, then, in a cylindrical Galilean frame $\{r, f, z, t\}$ which is rotating with the cylinder about its $z$-axis, there is an *induced* substratum flow $\mathbf{w} = -w\,r\,\hat{\mathbf{e}}_f$. Using the form of the tensor $\mathbf{grad}\,\mathbf{w}$ in cylindrical coordinates [15], we find that

$$\mathbf{grad}\,\mathbf{w} = w\,\hat{\mathbf{e}}_f\,\hat{\mathbf{e}}_r - w\,\hat{\mathbf{e}}_r\,\hat{\mathbf{e}}_f \quad . \tag{5-7a}$$

For a test body at rest on the inner surface of the cylinder, $\mathbf{v} = \mathbf{0}$, $\mathbf{a} = \mathbf{0}$, and $d\mathbf{w}/dt = \mathbf{v} \cdot \mathbf{grad}\,\mathbf{w} + \partial_t \mathbf{w} = \mathbf{0}$. Hence, $\mathbf{u} = -\mathbf{w}$ and $d\mathbf{u}/dt = \mathbf{0}$. Thus, by (5-4), there is the ponderable centripetal force,

$$\mathbf{F}_{ponderable} = -g\,m\,w^2\,r\,\hat{\mathbf{e}}_r = -(1/\sqrt{1 - w^2\,r^2/c^2}\,)\,m\,w^2\,r\,\hat{\mathbf{e}}_r \quad , \tag{5-7b}$$

acting on the body. The relativistic inertial factor is again $g\,m$ .

If we reverse the direction of the flow by spinning the cylinder in the opposite direction, we will obviously obtain exactly the same force. Whereas this independence of the physics on the *direction* of the flow was not so obvious in the gravitational case, it is familiar from everyday experience in the rotational case.

The three simple examples we have given are *canonical*, because they can be used to point out the essential distinctive characteristics of gravitational flows, inertial flows, and non-inertial flows. For the purpose of clarity in our discussion, we will call the vector **grad** $w^2$ the *inhomogeneity vector* of the flow **w** . It is independent of the direction of the flow, and it points in the direction of greatest increase in the *speed w* of the flow.

- A purely *gravitational flow* **w** is irrotational (**curl w** = **0**), and its inhomogeneity vector is *parallel* to the flow. In *Example 1* above, the gravitational inflow and outflow solutions both have the same inhomogeneity vector, and so they give rise to the same ponderable gravitational force.

- An *inertial flow* **w** is temporally constant and spatially homogeneous. Thus, its inhomogeneity vector is everywhere zero (these are the flows of Special Relativity).

- A purely *non-inertial flow* **w** is solenoidal ($div\,$**w** = 0), and its inhomogeneity vector is *perpendicular* to the flow. In *Example 3* above, the clockwise and counterclockwise rotations have the same inhomogeneity vector, and so they produce the same ponderable centripetal force.

## 6. Celestial Mechanics of Artificial Satellites

The fundamental distinctions between a celestial mechanics which is based on Newtonian *action-at-a-distance* gravitational theory and a celestial mechanics which is based on a theory of *substratum flow* are made evident in a study of the non-relativistic (slow motion) approximation to the fully relativistic geodesic equations of motion (4-16). The *non-relativistic* approximation of the relativistic equation in the form (4-15) is

$$\frac{d\mathbf{u}}{dt} = -(\mathbf{grad\,w}) \cdot \mathbf{u} \quad . \tag{6-1}$$

In terms of **v** and **w** , this is

$$\frac{d\mathbf{v}}{dt} - \frac{d\mathbf{w}}{dt} = -(\mathbf{grad\,w}) \cdot (\mathbf{v} - \mathbf{w}) \quad , \tag{6-2}$$

where $d\mathbf{w}/dt$ is the derivative of the parametric vector $\mathbf{w}(t) = \mathbf{w}(\mathbf{r}(t), t)$ along the path $\mathbf{r}(t)$ specified by the velocity vector $\mathbf{v} = d\mathbf{r}/dt$ . By the chain rule of differentiation,

$$\frac{d\mathbf{w}}{dt} \;=\; \mathbf{v}\cdot\mathbf{grad\,w} + \partial_t\mathbf{w} \quad . \tag{6-3}$$

Using the vector identities

$$\mathbf{v}\cdot\mathbf{grad\,w} - (\mathbf{grad\,w})\cdot\mathbf{v} = (\mathbf{curl\ w})\times\mathbf{v} \tag{6-4}$$

and

$$(\mathbf{grad\,w})\cdot\mathbf{w} = \frac{1}{2}\,\mathbf{grad}\,w^2 \quad , \tag{6-5}$$

we find that

$$\frac{d\mathbf{v}}{dt} \;=\; \frac{1}{2}\,\mathbf{grad}\,w^2 \;+\; (\mathbf{curl\,w})\times\mathbf{v} \;+\; \partial_t\mathbf{w} \quad . \tag{6-6}$$

To remind ourselves that the Galilean acceleration $\mathbf{a} \equiv d\mathbf{v}/dt$ is that of a *free* test body (think of a satellite), we write this as

$$\boxed{\;\mathbf{a}_{free} \;=\; \frac{1}{2}\,\mathbf{grad}\,w^2 \;+\; (\mathbf{curl\,w})\times\mathbf{v} \;+\; \partial_t\mathbf{w}\;} \quad . \tag{6-7}$$

The first term on the right-hand side of this equation represents the gravitational and generalized centrifugal accelerations, the second term represents the generalized Coriolis acceleration, and the third term is the acceleration arising from an explicit dependence of the flow on time.

This equation provides us with an easy way to study the problem of motion of test bodies in rotating and accelerated frames of reference in ordinary Newtonian mechanics. Ordinary Newtonian mechanics assumes that the underlying substratum is that of absolute immutable space. In our flow picture, this amounts to making the assumption that $\mathbf{w} = \mathbf{0}$ everywhere in some special non-rotating and non-accelerating Galilean frame. With this substratum of ordinary mechanics, we can study the problem of motion of test bodies by focusing on the flows that are *induced* by the motion of the rotating and accelerating frames. In other words, in the rotating and accelerating frames of ordinary Newtonian mechanics, absolute space will appear to be flowing because of the relative motions of the frames and absolute space. As an example, a rotating Galilean frame with a time dependent axial rotational velocity $\boldsymbol{\omega} \equiv \boldsymbol{\omega}(t)$ will induce the flow

$$\mathbf{w}(\mathbf{r},t) \ = \ \mathbf{r} \times \boldsymbol{\omega}(t) \ . \tag{6-8}$$

This flow can be substituted into equation (6-7) to learn everything one needs to know about the effects of the rotational motion of the Galilean frame on the motion of test bodies as they are observed in the rotating Galilean frame.

There is an entire realm of non-relativistic mechanics available for exploration *beyond* the realm of ordinary Newtonian mechanics when one allows the substratum to flow in complicated ways. Equation (6-7) holds for *arbitrary* flows of physical space and not just for the flows induced by the motion of frames in absolute immutable space. The canonical example of a flow that is not induced by the motion of a frame is the flow associated with an *isolated* spherically symmetric gravitational attractor. As we mentioned in Section 5, in a spherical Galilean frame centered on the attractor, this is given by one or the other of the flows

$$\mathbf{w}(\mathbf{r}) = \pm\sqrt{2GM / r} \ \hat{\mathbf{e}}_r \ , \tag{6-9}$$

where $G$ is the gravitational constant and $M$ is the mass of the attractor (see also references [5] and [6]). The connection between these possible flow solutions and the ordinary Newtonian gravitational theory is made through the heuristic equation,

$$w^2 \ = \ const. \ - \ 2\boldsymbol{\psi} \ , \tag{6-10}$$

which we call the *Bernoulli analog*. This relates the gravitational potential $\boldsymbol{\psi}$ of ordinary Newtonian gravitational theory to the square of the speed of the flow in the approximately corresponding spatial flow version. It has to be approximately correct, because ordinary Newtonian gravitational theory is approximately correct. When we substitute (6-10) into equation (6-7), we get $\mathbf{a}_{free} = -\mathbf{grad}\,\boldsymbol{\psi}$ for a time-independent gravitational flow, and this is the basic equation of ordinary Newtonian gravitational theory. The constant in (6-10) is determined by the boundary conditions for the speed of the flow *near the sources*. In other words, when there are several point sources present, the boundary conditions are that the speed of the flow near each source of strength $M$ must approach $\sqrt{2GM/r}$ . In the case of a single attractor, the constant is zero, and $y = -GM/r$ . In general, the constant will vary from streamline to streamline. The Bernoulli analog can also be used to elucidate the connection between rotational flows and the centrifugal potential.

In calling (6-10) the Bernoulli analog, we do not intend to suggest that the flow of the substratum obeys Euler's equation as it is normally derived by applying Newton's Second Law to a perfect fluid having material density (refer to any textbook on Fluid Mechanics). On the contrary, we are attempting to *elucidate* the generalization of Newton's Second Law by means of a flowing substratum. The substratum must not be given inertia, because this is what it is being used to *explain*.

In this paper, we are not making any assertions about what are the appropriate equations of motion for the substratum, *itself*. Within the context of General Relativity, one can proceed directly to equations of motion for **w** by substituting the flow metric (3-4) into Einstein's equations,

$$R_{nn} - \frac{1}{2} g_{nn} R = \frac{8pG}{c^4} T_{nn} \quad .$$

(6-11)

However, a discussion of the correct equations of motion for the substratum must be reserved for a future publication.

Without going into great detail here, we can *educe* that the flow in the simple non-rotating two-body problem will be approximately along the usual gravitational field lines which are obtained by adding the Newtonian potentials of the two bodies. When we apply the Bernoulli analog to the superposition of the potentials and also apply the appropriate boundary conditions for the flow speeds *near the sources*, we find that the flow structure divides into two separate flows, each separate flow being associated with one of the two bodies. As a result, there will be a boundary or a *surface of transition* separating the flows. This is schematically represented in figure (6-12) by the dotted line. The speed of the flow is schematically represented by the density of the streamlines.

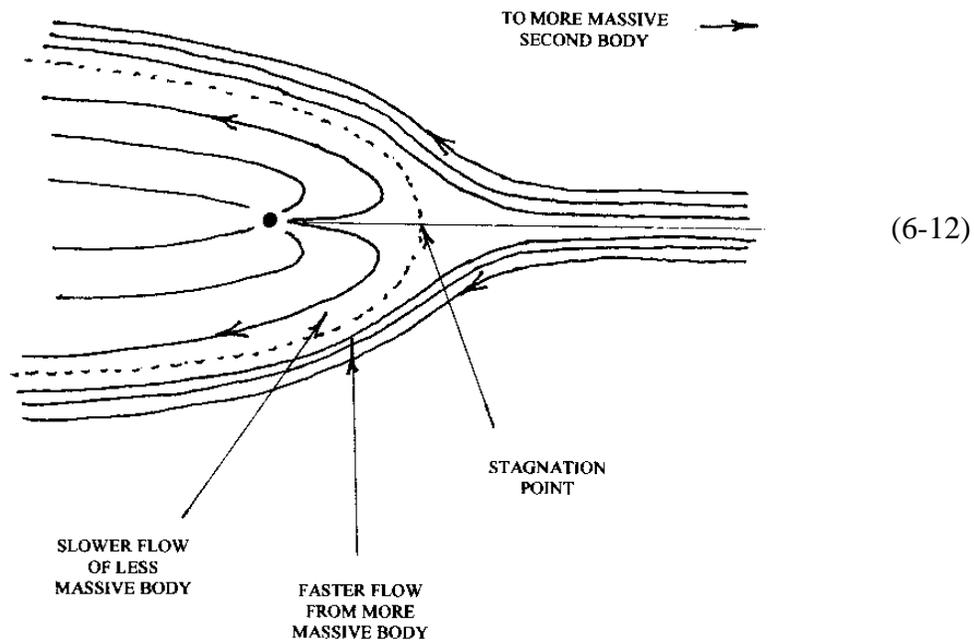

(6-12)

STAGNATION
POINT

SLOWER FLOW
OF LESS
MASSIVE BODY

FASTER FLOW
FROM MORE
MASSIVE BODY

When the less massive body is orbiting the more massive body, it is possible that the stagnation point is turned somewhat in the direction of the orbit. In this two-body solution, there will be a significant speeding up of an atomic clock at the stagnation point ($\mathbf{w} = \mathbf{0}$) (see the time dilation equation (3-6)). This is quite different from what one would normally expect in the non-flowing two-body solution, and it is the basis of the satellite experiment which was suggested in reference [5]. There will also be time dilation jumps for a freely running clock when the satellite crosses the surface of transition  anywhere in the general region shown in figure (6-12).

What happens to the flow along the surface of transition really needs to be elucidated by experiment. There may be a strict discontinuity between the separate flows, there may be turbulent flow there, or there might be a smooth transition between the flows (in which case, there will be non-vanishing curl in the transition surface). In the case of a smooth transition, the $(1/2)\mathbf{grad}\,w^2$ term on the right hand side of equation (6-7) tells us there will be a forward acceleration of a test satellite as it passes through the transition surface from the region of slower flow into the region of faster flow (this is shown schematically in figure (6-13)). In certain two-body configurations, this might provide a substitute for propulsion. It is a kind of slingshot effect. Contrariwise, motion from the region of faster flow into the region of slower flow will cause a deceleration.

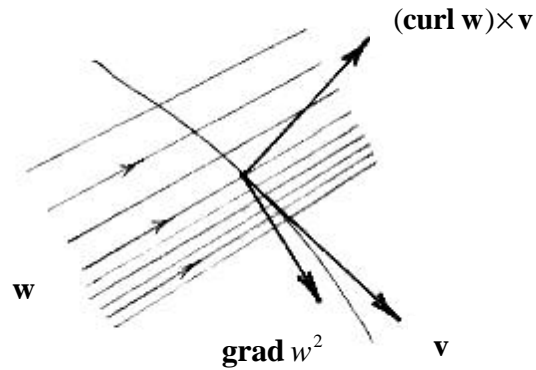

(6-13)

    As is also shown schematically in figure (6-13), the generalized Coriolis acceleration (**curl w**)×**v** of equation (6-7) will act on a satellite moving from a region of slower flow to a region of faster flow so as to reveal the general direction of the flow. In the case when **v** is perpendicular to the flow **w**, the Coriolis acceleration will be exactly in the direction of flow. This is the basis of the satellite experiment which was suggested in reference [6].

## Acknowledgement

This publication was made possible by a
gift from Harold and Helen McMaster.